\documentclass[12pt]{article}
\usepackage{helvet,times,mathptm}
\usepackage{graphicx}
\newcommand\beq{\begin{equation}}
\newcommand\eeq{\end{equation}}
\newcommand\bea{\begin{eqnarray}}
\newcommand\eea{\end{eqnarray}}

\begin{document}

\begin{center}
{\noindent
{\bf Fully Symmetrized VB Based Technique for Solving Exchange Hamiltonians 
of Molecular Magnets}}
\end{center}

\begin{center}
{\noindent

{\small \bf Shaon Sahoo$^{a,}$ \footnote[1]{shaon@physics.iisc.ernet.in}, 
Raghunathan Rajamani$^{b,}$ \footnote[2]{rajamani@sscu.iisc.ernet.in}, 
S. Ramasesha$^{b,}$ \footnote[3]{ramasesh@sscu.iisc.ernet.in}}
and {\small \bf Diptiman Sen$^{c,}$} \footnote[4]{diptiman@cts.iisc.ernet.in}}

\end{center}
{\noindent 
{\small 
a. Department of Physics, Indian Institute of Science, Bangalore 560012, India.
\\
b. Solid State $\&$ Structural Chemistry Unit, Indian Institute of Science, 
Bangalore 560012, India.\\
c. Centre for High Energy Physics, Indian Institute of Science, 
Bangalore 560012, India.
\vspace{.2cm}
~\\
}}
{\noindent
{\bf Abstract}
Generally, the first step in modeling molecular magnets involves obtaining the 
low-lying eigenstates of a Heisenberg exchange Hamiltonian which conserves total
spin and belongs usually to a non-Abelian point group. In quantum chemistry, it 
has been a long standing problem to target a state which has definite total 
spin and also belongs to a definite irreducible representation of the point 
group. Many attempts have been made over years, but unfortunately these have 
not resulted in methods that are easy to implement, or even applicable 
to all point groups. Here we present a general technique which is a hybrid 
method based on Valence Bond basis and the basis of z-component of the 
total spin, which is applicable to all types of point groups 
and is easy to implement on computer. We illustrate the 
power of the method by applying it to the molecular magnetic system, 
$Cu_6Fe_8$, with cubic symmetry. We emphasize that our method is applicable 
to spin clusters with arbitrary site spins and is easily extended to 
fermionic systems.}
 
\section{Introduction}
The field of molecular magnetism has witnessed an explosion in the number of 
systems that exhibit molecular magnetic phenomena such as quantum resonant 
tunneling and photomagnetism (see reviews \cite{gat1,gat2}). This explosive 
growth has also presented 
challenges to theorists modeling these systems. The problems encountered by 
theorists begin with modeling the nature of exchange interactions between 
pairs of magnetic ions. While an electronic many-body Hamiltonian has to be 
solved for determining the nature of exchange, this is often circumvented 
by guessing the nature of exchange based on the knowledge of the ligands, 
electron configuration of the transition metal ion and the geometry of 
the complex. The second problem concerns with obtaining the eigenstates 
of the exchange Hamiltonian,

\begin{eqnarray}
\label{eqn1}
H_{ex}=-\sum_{\langle ij \rangle}J_{ij}\hat{S_i}
\cdot\hat{S_j}
\end{eqnarray}

\noindent describing the coupling between pairs of magnetic 
ions $\langle ij \rangle$ with exchange constant $J_{ij}$. Often, the Fock 
space of the Hamiltonian of the magnetic system could 
be very large (in case of $Mn_{12}Ac$, it is as large as a hundred million 
\cite{raghu} and  in $Fe_{12}$ ferric wheel, it is more than two billion 
\cite{indra}) and obtaining even a few low-lying states of the Hamiltonian 
could pose a challenge.  Since the exchange Hamiltonian 
conserves both total spin and z-component of total spin ($M_S$), the 
problem can be simplified by specializing the basis, in which the 
matrix representation of the Hamiltonian is sought, to the case of 
fixed z-component of the total spin. Further simplification 
could come from exploiting spatial symmetries of the model. An ideal situation 
would correspond to one in which all the spin and spatial symmetries are 
utilized to construct a fully symmetrized basis to minimize the size of 
the Hamiltonian matrix that needs to be diagonalized. 

The conservation of the $S_{tot}^{z}$, the total z-component of 
spin is trivially achieved by choosing from the Fock space, states whose total 
$M_{S}$ value corresponds to the desired value. This is possible since 
individual $S_{i}^{z}$ operators commute with the $S_{tot}^{z}$ operator. 
It is also quite straightforward to set up the Hamiltonian matrix in this 
basis and solve for a few low-lying states in cases where the Hilbert space 
is spanned by a few hundred million states \cite{pati}. However, the exchange 
Hamiltonian of molecular magnets often consists of interactions that 
are geometrically frustrated.  In such a system, the ground 
state spin is often not predictable and one needs to obtain 
the lowest state in each total spin subspace to fix the spin of 
the ground state. Besides, low energy states with different total 
spins lie close in energy and it is numerically difficult to achieve 
convergence to nearly degenerate eigenstates unless they can be dispersed into 
orthogonal Hilbert spaces. We can partially alleviate this problem by employing 
the parity symmetry of the exchange Hamiltonian. This symmetry corresponds to 
rotation of all the spins in the system around the x or y-axis by an angle 
$\pi$ which leaves the Hamiltonian invariant, in the $M_{S}=0$ sector. 

\begin{eqnarray}
\left[H,R_y(\pi)\right] = 0; {~\rm where, ~}
R_y(\theta) = e^{-i\theta\hat{S}_y/\hbar} 
\end{eqnarray}

\noindent The action of parity operator ($\hat{P}$) on a basis state with site 
$m_s$ values $m_1$, $m_2$, $m_3$ .... $m_n$, is to flip all 
the spins in the system, i.e., 

\begin{eqnarray}
\hat{P}|m_1 m_2 \ldots m_n\rangle = (-1)^{\eta}|-m_1 -m_2 
\ldots-m_n\rangle; {~\rm where, ~}  \eta = \sum_i s_i
\end{eqnarray}

Thus, the parity operator which conserves the total $M_{S}$ value, 
only when $M_{S} = 0$, is a symmetry element of the Hamiltonian matrix  in 
the $M_{S} = 0$ sector. In the general case where the individual objects have 
spin $s_{i}$, if $\sum_{i}s_{i}$ is even then symmetric (antisymmetric) 
combination of the basis states, under parity, will span a space of even (odd) 
total spin states. The method can be extended easily to systems which allow 
only half odd-integer total spin. Since in most cases, 
the lowest excited state usually has a spin which is one different from that 
of the ground state, this symmetry renders it easy to obtain the spin gap 
accurately. However, the size of the Hilbert space is only reduced by 
approximately half of the size of the full $M_{S} = 0$ space, by using this 
symmetry. Use of parity is still advantageous as exploiting spatial 
symmetries is straightforward, even when the point group is not Abelian.

Construction of spin adapted configuration state functions (CSF) 
has been a problem of long standing interest in quantum chemistry. The CSFs are 
simultaneous eigenstates of total $S_{tot}^2$ and $S_{tot}^z$ and setting up 
the Hamiltonian matrix in this basis leads to matrices of smaller size besides 
allowing automatic labeling of the states 
by the total spin. Besides, the eigenvalue spectrum is enriched, since we can 
obtain several low-lying states in each total spin sector. This is in contrast 
to obtaining several low-lying states in a given total M$_S$ sector as the 
latter would have states with total spin S$_{tot}\geq$M$_S$. There are 
many ways of achieving this task \cite{pauncz}. The simplest method involves 
setting up the matrix of the total spin operator, $S_{tot}^{2}$, in the CSF 
basis of fixed $M_{S}$ and obtain the eigenstates corresponding to given total 
spin value; these eigenstates which are linear combinations of the constant 
$M_{S}$ CSFs are then the spin adapted CSFs. For large systems this method is 
not practical. Another method which is some times used is the 
L$\ddot{o}$wdin projection method \cite{lowdin,bernu} in which a projection 
operator $P_S$, given by,

\begin{eqnarray}
P_{S}=\Pi_{S'\ne S}(\hat S^{2}-S'(S'+1)) 
\end{eqnarray}

\noindent is used to project out all undesired states of spins ${S}'$ 
except the spin S of interest, from a given CSF. The methods that 
have been extensively in vogue for construction of the 
spin adapted CSFs are the Graphical Unitary Group Approach (GUGA) 
\cite{saxe,sie,brooks}, Symmetry Group Graphical Approach (SGGA) \cite{duch} 
and the Valence Bond (VB) approach \cite{pauling,eyring,soos,ramasesh,mazum}. 
In the GUGA method, the total spin adapted CSFs are represented as Shavitt 
graphs or Paldus arrays and the matrix element of a term (which corresponds 
to a generator of the unitary group) in the Hamiltonian between any two CSFs 
is obtained by comparing the two arrays or graphs corresponding to the CSFs. 
Similarly, graphs are used to represent spin adapted CSFs in the SGGA method 
and rules for computing matrix elements between two CSFs of a term in the 
Hamiltonian have been derived (see for example 
reviews \cite{duch1,duch2,robb}). 
Using these methods it is possible to carry out large scale configuration 
interaction (CI) calculations. While in all these methods the total spin 
symmetry is fully exploited, spatial symmetry adaptation is not an easy task 
\cite{duch1}. The CSFs are each built up of several orbitals with each orbital 
in general transforming according to some specific irreducible representation 
of the point group of the system.  The direct product of the irreducible 
representations of a general symmetry group is not a single irreducible 
representation of the same group. Thus it is not possible to associate an 
irreducible representation with a given CSF unless the point group to which 
the system belongs is an Abelian group \cite{shavitt,shavitt1}. Otherwise, 
symmetry operation on a CSF leads to a linear combination of many CSFs which 
is in general difficult to construct. For small dimensionalities of the Hilbert 
spaces, matrix representation of the symmetry operators can be obtained in the 
space of CSFs. The projection matrix for a given irreducible representation 
can be constructed from these matrices and from these, the symmetry adapted 
CSFs \cite{bue}. However, this approach is of limited value in real large 
scale problems \cite{duch1}. In quantum chemical literature this difficulty 
is bypassed by dealing with Abelian subgroups of the system’s point group 
\cite{ret,sha}. But, this can lead to ambiguities in assigning the irreducible 
representation of a state \cite{ret1}. Among the methods for constructing 
spin adapted CSFs, the VB method is the simplest and will be considered in 
the next section. 

The ultimate goal of symmetry adaptation is to be able to 
exploit the full spatial and spin symmetries of the system, both for 
computational efficiency and for labeling of the state by the irreducible 
representation to which it belongs. In the next section, we give a brief 
introduction to the symmetrized valence bond (VB) approach that was developed 
earlier and highlight the difficulties associated with the technique 
\cite{ramasesh1}. In the third section, we present a hybrid VB-constant $M_S$ 
method which overcomes these difficulties. In the fourth section, we 
illustrate an application of this method to a magnetic spin cluster. In 
the final section, we summarize and discuss the future prospects for 
the technique.

\section{The Symmetrized VB Approach} 
Exploiting the invariance of both total spin and its z-component is 
nontrivial, since eigenstates of the $S^{z}_{tot}$ operator expressed as a 
product of the eigenstates of all the $S_{i}^{z}$ operators are not 
simultaneously eigenstates of the $S^{2}_{tot}$ operator. The situation 
is further complicated by the fact that in a molecular magnet, often the 
spins of all the constituent magnetic centers, $s_{i}$ are not the same. In 
such a situation, the easiest way of constructing the spin adapted functions 
is the diagrammatic valence bond (VB) method based on modified Rumer-Pauling 
rules \cite{pauling,soos}. In this method, a magnetic site with a given spin 
$``s_{i}"$ is replaced by $2s_{i}$ spin-half objects. To obtain a state 
with total spin $S$ from N such spin-1/2 objects from all the magnetic 
centers, ($N-2S$) of these spin-1/2 objects are singlet spin paired 
explicitly, subject to the following restrictions: (1) there should be no 
singlet pairing of any two spin-half objects belonging to the same magnetic 
center (this ensures that the $2s_{i}$ objects are in a totally symmetric 
combination \cite{arovas}) , (2) a total of $2S$ spin-half objects are 
left unpaired, (3) when all the spin-half objects are arranged at the vertices 
of a regular polygon with number of vertices equal to number of spin-half 
objects, N, and straight lines are drawn between spin paired vertices, there 
should be no intersecting lines in the resulting diagram and (4) when all the 
spin-half objects are arranged on a straight line and lines are drawn between 
spin paired objects, these lines should not enclose any unpaired spin-1/2 
object. These rules follow from the generalization of the Rumer-Pauling 
rules to objects with spin greater than 1/2 and total spin greater than zero. 
The set of diagrams which obey these rules would hence forth be called 
$``$legal$"$ VB diagrams. Some legal VB diagrams are shown in Fig. 1. A line 
in the VB diagram between two spin -1/2 objects i and j corresponds to the 
state ${(\alpha_{i} \beta_{j}-\beta_{i} \alpha_{j})}$/$\sqrt{2}$ where we 
choose $\alpha$ to correspond to $|\uparrow \rangle$ and $\beta$ to 
$|\downarrow \rangle$ orientations of the spin. The phase 
convention assumed is that the ordinal number $``i"$ is less than the ordinal 
number $``j"$. The $2S$ spin-1/2 objects $k_{1}$ $k_{2}$ 
$k_{3}$ . . . . $k_{2S}$ which are left unpaired 
can be taken to represent the state with $M_S = S$ given 
by $\alpha_{k_1} \alpha_{k_2} \alpha_{k_3} . . . \alpha_{k_{2s}}$. VB states 
corresponding to other M$_S$ value for this state with spin S, can be obtained 
by operating the S$_{tot}^{-}$ operator on the state by the required number of 
times. Since the exchange Hamiltonian is isotropic, each eigenstate in the 
spin S is (2S+1) fold degenerate and it is sufficient to work in subspace of 
chosen M$_S$ value. The VB 
state corresponding to a given diagram is a product of the states representing 
the constituent parts of the diagram. On a computer, a $``$legal$"$ VB diagram 
of any spin can be uniquely represented by an integer of 2N bits with a one 
bit representing the beginning of a singlet line and a zero bit the ending of 
singlet line. The unpaired spins are also represented as one-bits. Fig. 1 
also shows bit representation of typical VB diagrams. 

\begin{figure}
\begin{center}
\hspace*{1cm}{\includegraphics[width=10.0cm]{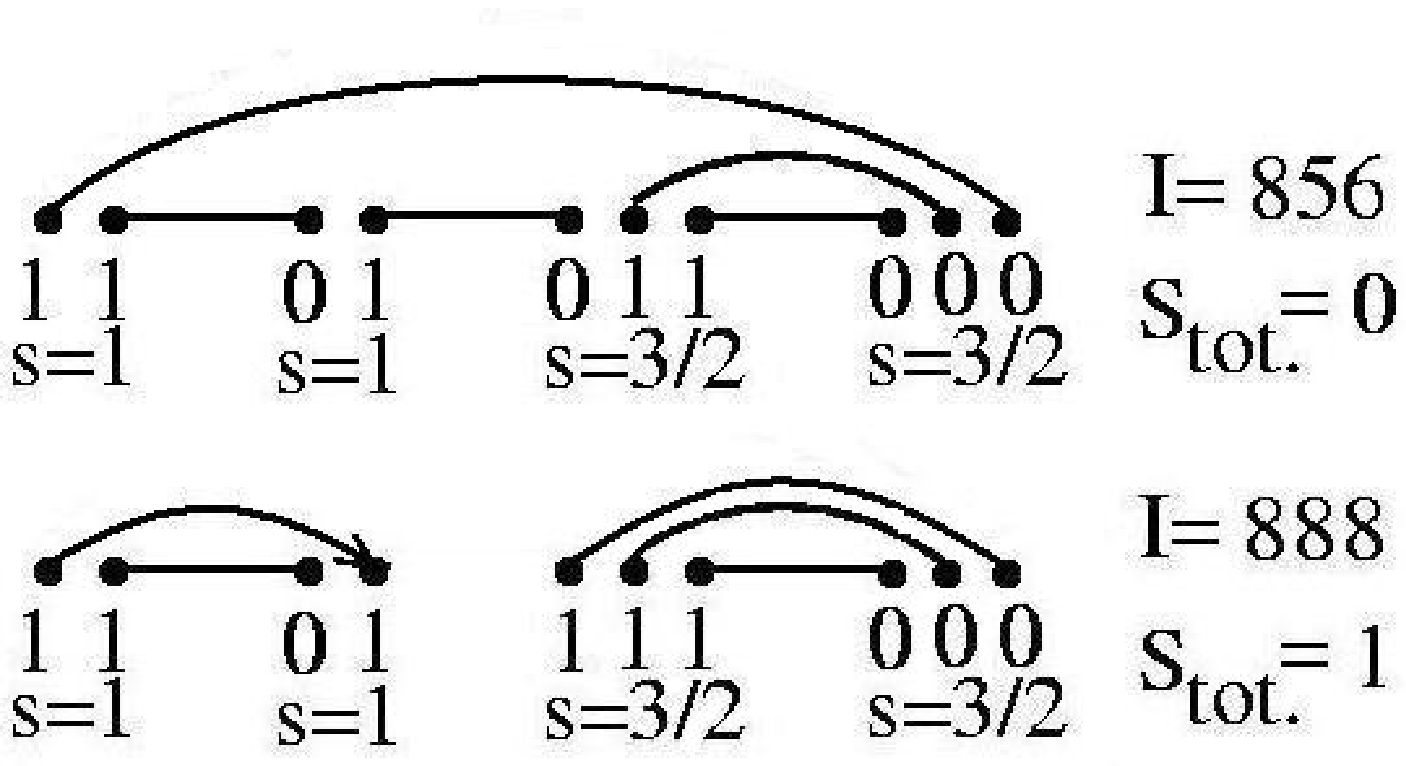}}
\caption{Above VB diagram shows spin pairings to yield a total spin 
$S_{tot}$=0 state from ten spin-1/2 objects, constituent elementary spins 
of two spin 1 and two spin 3/2. Its bit representation corresponds to unique 
integer I = 856. The bottom VB diagram shows a $S_{tot}$=1 state, 
the corresponding unique integer is I = 888.}
\end{center}
\end{figure}

To spatially symmetrize a VB basis, it is necessary to know the 
result of a symmetry operator operating on a legal VB diagram. In general, 
the resultant of such an operation on a $``$legal$"$ VB diagram is an 
$``$illegal$"$ VB diagram. An example of this is shown in Fig. 2 
\cite{ramasesh1}. An illegal VB diagram can be decomposed into a linear 
combination of legal VB diagrams, a process that is computationally 
demanding. In practice, the VB space is broken down into smaller invariant 
subspaces of the symmetry group and within each invariant space, a 
symmetrized linear combination of the VB basis is constructed. 
However, the structure of the invariant spaces is very complex 
and constructing disjoint invariant spaces is not simple. While constructing 
symmetrized VB basis, a projection operator for a given symmetry 
representation is employed to project the symmetrized linear combinations by 
acting on each of the VB states in the invariant space. While the number of 
linearly independent symmetry combinations of a given representation is known 
{\it a priori}, the actual linear combinations are obtained by carrying out 
Gram-Schmidt orthonormalization of the projected states.  However, since the 
VB diagrams are not orthogonal the orthonormalization process is both 
computationally involved and time consuming.

\begin{figure}
\begin{center}
\hspace*{1cm}{\includegraphics[width=12.0cm]{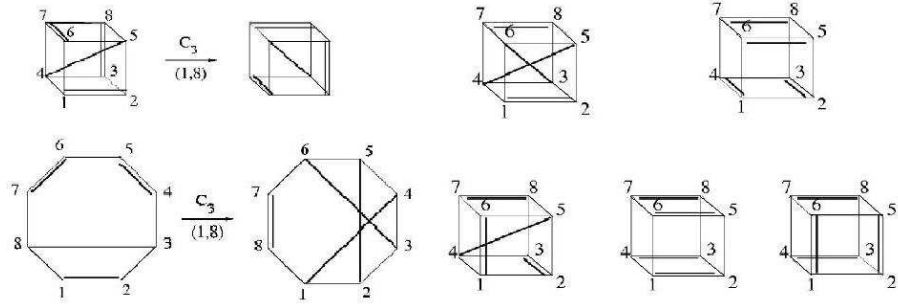}}
\caption{The effect of operation by the C$_{3}^{1}$ symmetry operator 
about the (1,8) axis. Top left shows the initial and final VB diagrams with 
spin couplings between vertices of the cube shown as dark lines. Bottom left 
shows the same states as spin couplings between vertices of a regular octagon. 
The resultant is an $``$illegal$"$ diagram. Decomposing the resultant to 
$``$legal$"$ VB diagrams yields a sum of five VB diagrams shown on the right, 
with spin couplings between vertices on a cube.}
\end{center}
\end{figure}

Furthermore, in case of molecular magnets containing magnetic 
ions with spin greater than half, the exchange operator between such high-spin 
centers also generates ``illegal" VB diagrams as it involves non-nearest 
neighbor exchange interactions between constituent elementary spins 
\cite{soos}. To illustrate, exchange between a center A with say spin one and 
a center B with spin 3/2, $S_{A} \cdot S_{B}$, is expressed as 
$(s_{A_1} + s_{A_2}) \cdot (s_{B_1}+s_{B_2}+s_{B_3})$. These exchange terms 
operate on a VB diagram with constituent elementary spins which are 
non-nearest neighbors and in general generate $``$illegal$"$ VB diagrams 
as per the VB rules. Decomposition of the resultant $``$illegal$"$ VB 
diagrams to $``$legal$"$ VB diagrams could present a serious bottle-neck in 
computing the Hamiltonian matrix elements. In view of these difficulties, 
a fully symmetrized VB approach to solving Heisenberg exchange Hamiltonian 
particularly in the context of  molecular magnets is not feasible.

\section{Hybrid method based on VB Basis and Constant M$_{S}$ Basis} 
In the constant $M_{S}$ basis, a basis state of an ensemble of spins $s_{1}$,
$s_{2}, \cdots$, $s_{N}$, is represented by a direct product of the $m_{s}$ 
states
of each spin such that the total $M_{S}=\sum m_{i}$. The basis states in the
constant $M_{S}$ representation are orthonormal by construction. Given the
definition of a line in the VB diagram, every VB diagram can be broken up
into a linear combination of the constant $M_{S}$ basis states. A VB diagram
with $p$ singlet lines give rise to $2^{p}$ basis states in the constant $M_{S}$
basis. To effect the conversion of VB diagrams to constant $M_{S}$ functions,
each singlet line gives two states; in one state, the site at which a singlet
line begins is replaced by an $\alpha$ spin and the one at which it ends by a
$\beta$ spin with phase +1 and in the other the spins are reversed and the
associated phase is -1. Once the VB diagrams are converted to constant $M_{S}$
basis states with constituent spins, it is possible to associate a $m_{i}$ 
value with a composite spin, given by 
$m_i = (n_{i\uparrow} - n_{i\downarrow})/2$, where $n_{i\uparrow}$ is the 
number of up-spin halves and $n_{i\downarrow}$ is the number of down-spin 
halves at the site i. However, there is a normalization constant w$_i$, which
follows from Clebsch-Gordan coefficients, given by,

\begin{eqnarray}
w_{i}=\left[{\frac{(2s_{i})!}{(s_i+m_{i})!(s_i-m_{i})!}}\right]^{-1/2} 
\end{eqnarray}

\noindent for a site with 
spin s$_{i}$ in state $m_{i}$ \cite{arovas}. We can assume without loss
of generality that the $M_{S}$ value of the VB diagram is also $S$.
Computationally, finding the transformation of a state in the VB basis to
constant $M_{S}$ basis is straightforward. We initialize the coefficients in
the constant $M_{S}$ basis to zero. We then decompose, sequentially, each VB
diagram into constant $M_{S}$ states and update the coefficient of the basis
state of corresponding $M_{S}$ by adding to it the VB coefficient
times the product of Clebsch-Gordan factors with appropriate phases. The
matrix relating the VB basis states to constant $M_{S}$ basis states,
{\bf C}, is a $V\times M$  matrix, where $V$ is the dimensionality of the VB
space and $M$ that of the constant $M_{S}$ space.

The construction of the projection operator for projecting all
the states of a given symmetry representation in a given spin space can now
be accomplished. We construct the matrix representation of a symmetry
operator, $\hat R$, of the point group in the chosen spin space by operating 
with $\hat R$ on
each state in the constant $M_{S}$ basis and searching for the resulting state
in the list of $M_{S}$ basis states. Each basis state in this representation is
carried over to another basis state by a symmetry operation of the point group.
Thus, the matrix {\bf R$_M$} though an $M \times M$ matrix contains only one nonzero 
element in each row; this makes manipulations with this matrix computationally
fast. The knowledge of the {\bf C} and the {\bf R$_M$} matrices give the
effect of operating by the symmetry operator $\hat R$ on a VB state as a
linear combination of the constant $M_S$ basis states via the matrix
{\bf B$_{R}$=CR$_M$}. The projection operator for projecting out the basis 
states on to a chosen irreducible representation of the point group $\Gamma$ 
is given by,

\begin{eqnarray}
P_{\Gamma}=\sum_{\hat R}\chi_{\Gamma}^{irr}({\hat R}){\hat R}
\end{eqnarray}

\noindent where, $\chi_{\Gamma}^{irr}(\hat R)$ is the character under the 
symmetry operation $\hat R$ in the character table of the point 
group of the system \cite{bishop}. The matrix representation of 
$P_{\Gamma}$ in the mixed VB and constant $M_S$ basis is given by,

\begin{eqnarray}
{\mathbf Q_{\Gamma}}=\sum_{R}\chi_{\Gamma}^{irr}(R){\mathbf B_R}
\end{eqnarray}

\noindent where, {\bf Q$_{\Gamma}$} is a $V \times M$ matrix. However, the 
rows of the matrix {\bf Q$_{\Gamma}$} are not linearly independent, 
since the symmetrized basis spans a much smaller dimensional Hilbert 
space. The exact dimension of 
the Hilbert space spanned by the system in the irreducible representation 
$\Gamma$ can be known {\it a priori}. The dimension of the space $\Gamma$, 
V$_{\Gamma}$, is given by,

\begin{eqnarray}
V_{\Gamma}=(d_\Gamma /h)
\sum_{\hat R}\chi(\hat R)\chi_{\Gamma}^{irr} (\hat R)
\end{eqnarray}

\noindent where $d_\Gamma$ is
the dimensionality of the irreducible representation $\Gamma$, $h$ is the
number of elements in the point group and $\chi(\hat R)$ is the reducible 
character for the operation $\hat R$. The $V_{\Gamma}\times M$ projection 
matrix, {\bf P$_{\Gamma}$} of
rank $V_{\Gamma}$ is obtained by Gram-Schmidt orthonormalization of the rows
of the matrix {\bf Q$_{\Gamma}$} until $V_{\Gamma}$ orthonormal rows are
obtained. These $V_{\Gamma}$ orthonormal rows represent the projection
matrix {\bf P$_{\Gamma}$}. The $M \times M$ Hamiltonian matrix, {\bf H$_{M}$} 
is constructed in the constant $M_{S}$ basis which is described 
elsewhere \cite{pati}. Since the basis states in this representation 
are orthonormal, we do not encounter the problem of $``$illegal$"$ 
states as with the VB representation. The $V_{\Gamma}\times V_{\Gamma}$ 
Hamiltonian matrix in the fully symmetrized
basis is given by {\bf P$_{\Gamma}$}{\bf H$_{M}$} $({\bf P}_{\Gamma})^{\dag}$
and one could use any of the well known full diagonalization routines to
obtain the full eigenspectrum or use Davidson algorithm to get a few low-lying
states of the symmetrized block Hamiltonian in the chosen spin and symmetry
subspace.

The above procedure does not lead to the smallest block of the 
Hamiltonian matrix, when the irreducible representation for the block is 
degenerate, such as the E, T or H representations. In such cases, it is 
advantageous to work with bases that transform according to one of the 
components of the irreducible representation. This can be achieved by choosing 
an axis of quantization and projecting out bases states of the irreducible 
representation which are diagonal about a rotation about the quantization 
axes. For example, in the case of the irreducible representation that 
transforms as T, we can choose one of the C$_{3}$ axes as a quantization axis 
and project the basis states of the irreducible representation using 
(I+C$_{3}^{1}$+C$_{3}^{-1}$) as the projection operator. This operator projects 
states that transform as the Y$_{1}^{0}$ component of the three fold 
degenerate irreducible tensor operator. Similarly, for the E representation, 
we could use any of the C$_{2}$ axis as a quantization axis and use the 
projection operator (I+C$_{2}^{1}$) to project out basis states that 
transform as one of its components. This is equivalent to projecting out the 
states which transform as a given row of the irreducible representation; the 
latter are not listed in standard group theoretical character tables. 

Computation of static properties such as spin densities and 
spin-spin correlation functions in the eigenstates of the Hamiltonian is 
rendered simple due to the orthogonality of the constant $M_{S}$ basis. The 
site spin operators such as the z-component of the spin are diagonal in this 
basis, while other operators such as the raising and lowering operators,  
though not diagonal in this basis, have a very simple matrix 
representation. In computing various properties, the procedure we follow is 
to express the eigenstates in the unsymmetrized constant $M_{S}$ basis 
and to compute the desired properties using this representation.

The additional steps involved in the hybrid VB-Constant M$_S$ method are (i) 
constructing the {\bf C} matrix, whose i$^{\rm th}$ row contains the 
coefficients of 
the constant M$_S$ functions appearing in the i$^{\rm th}$ VB basis function. 
This is a very fast step as the constant M$_S$ states are an ordered 
sequence of integers and a VB state with $n$ lines is a linear 
combination of 2$^n$ constant M$_S$ functions. (ii) In the hybrid 
approach, computation of the {\bf C} matrix involves the matrix 
multiplication, {\bf C}($\sum_R\chi_{\Gamma}^{irr}(R)${\bf R$_M$}) = 
{\bf CR$_M'$}. The number of arithmetic operations involved 
is however very small, since both {\bf C} and {\bf R$_M'$} are sparse 
matrices. In both constant M$_S$ and hybrid approaches one has to 
obtain the projection operator {\bf P$_{\Gamma}$} by retaining only the 
orthogonal rows of the matrix {\bf R$_M'$} or {\bf Q$_{\Gamma}$} respectively. 
Since the number of orthogonal rows in {\bf Q$_{\Gamma}$} is far fewer than in 
{\bf R$_M'$}, this step is faster in the hybrid approach than in the constant 
$M_S$ approach by a factor D($\Gamma_S$)/D($\Gamma_{M_S}$), where  
D($\Gamma_S$) is the dimensionality of the space of the irreducible 
representation $\Gamma$ with spin S and D($\Gamma_{M_S}$) is similarly 
the dimension of the space $\Gamma$ with constant M$_S$. This advantage is 
largely off-set by the fact that the {\bf R$_M'$} matrix in constant 
$M_S$ basis is more sparse than the {\bf Q$_{\Gamma}$} matrix in the hybrid 
approach.Computation of the eigenvalues in the constant $M_S$ approach is 
slower than in the hybrid approach, since D($\Gamma_{M_S}$)$>$D($\Gamma_S$) 
for most S. This advantage may be partially off-set because the 
Hamiltonian matrix in the hybrid approach is usually more dense. 
The memory required for the hybrid approach is not very different from that 
of constant $M_S$ approach since the matrices though smaller in the hybrid 
approach, are slightly denser. The only additional array required in the 
hybrid approach is the storage of {\bf C} matrix, which is very sparse.
The major advantage of the hybrid approach is that we can obtain a far richer 
spectrum, since we are targeting each spin sector separately, unlike in the 
constant $M_S$ approach. Thus, if we can obtain, say 10 states in each S sector 
of the $2n$ spin-1/2 problem, we would have $10(n+1)$ unique states 
compared to the constant $M_S$ technique where many of these spin 
states would be repeated in different $M_S$ sectors.

\section{Application to $Cu_{6}Fe_{8}$} 
We have applied the above method to model the susceptibility behavior of the 
molecule [$(Tp)_{8}$ $(H_{2}O)_{6}Cu_{6}^{II}Fe_{8}^{III}(CN)_{24}$]
$(ClO_{4})_{4}\cdot 12H_{2}O\cdot 2Et_{2}O$ \cite{shi}, where Tp stands 
for hydrotris (pyrazolyl) borate (Fig. 3).
\begin{figure}
\begin{center}
\hspace*{1cm}{\includegraphics[width=7.0cm]{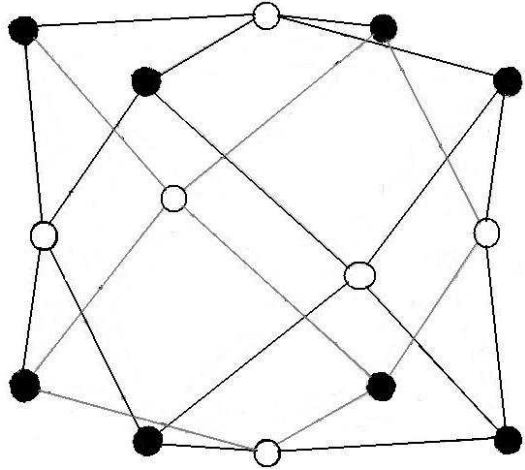}}
\caption{Schematic of $Cu_6Fe_8$ cluster. Filled and open circles correspond 
to Fe and Cu (both spin-1/2) sites respectively. Lines represent the 
exchange coupling between various spin sites.}
\end{center}
\end{figure}
In this molecule the $Cu^{II}$ ions as well 
as the $Fe^{III}$ are in spin-1/2 state. The eight $Fe^{III}$ ions are at cube 
corners and the six $Cu^{II}$ ions are on the outward perpendicular to the 
face centers of the cube. Each  $Cu^{II}$ ion is connected to the four nearest 
$Fe^{III}$ ions via ferromagnetic exchange interactions. There are no Fe-Fe or 
Cu-Cu interactions. 
The molecule has a spin 7 ground state. This system has a very high symmetry 
of the cube and incorporates all the complexities that can be encountered 
in the application of our technique. From the susceptibility data, the 
strength of the exchange interaction J, was estimated to be 30 cm$^{-1}$ 
\cite{shi}. Because of the high symmetry of this 
system, we chose this as an example for applying our technique. The 
dimensions of the various subspaces are given in Table 1.

\begin{table}
\caption{Dimensionalities of the total spin spaces of a system
of 14 spin-1/2 objects. $D(S)$ is the dimensionality of the constant S basis
and $D(M_S)$ is the dimensionality of the constant $M_S$ basis.}
\begin{center}
\begin{tabular}{|l||c|c|c|c|c|c|c|r|}
\hline
S/M$_{S}$ & 0 & 1 & 2 & 3 & 4 & 5 & 6 & 7 \\ \hline
$D(S)$ & 429 & 1001 & 1001 & 637 & 273 & 77 & 13 & 1
\\ \hline
$D(M_{S})$ & 3432 & 3003 & 2002 & 1001 & 364 & 91 &
14 & 1 \\
\hline

\end{tabular}
\end{center}
\end{table}

Using the hybrid VB-constant $M_{S}$ method, we have broken down the space in 
each total spin sector into basis states that transform as different 
irreducible representations of the cubic point group. The dimensionalities 
of the various subspaces are shown in Table 2. The subspaces transforming as 
the E representations are broken down into subspaces of half their dimension 
by quantizing the system along a C$_{2}$ axis. Similarly, the subspaces 
transforming as the T representations are broken down into one third their 
dimensions in Table 2, by using a C$_{3}$ axis as the axis of quantization. 

\begin{table}
\caption{Dimensionalities of various subspaces of 
the $Cu_{6}^{II}Fe_{8}^{III}$ cluster for irreducible representation $\Gamma$.}

\begin{center}
\begin{tabular}{|c||c|c|c|c|c|c|c|c|}
\hline
S$_{tot} \rightarrow$ &  &  &  &  &  &  &  &  \\ 
$\Gamma$ $\downarrow$ & 0 & 1 & 2 & 3 & 4 & 5 & 6 & 7 
\\ \hline \hline
A$_{1g}$ & 6 & 32 & 24 & 24 & 9 & 5 & 1 & 1 \\ \hline
A$_{2g}$ & 13 & 15 & 19 & 8 & 5 & 0 & 0 & 0 \\ \hline
E$_{g}$ & 34 & 90 & 90 & 60 & 26 & 10 & 2 & 0 \\ \hline
T$_{1g}$ & 78 & 165 & 171 & 99 & 39 & 6 & 0 & 0 \\ \hline
T$_{2g}$ & 66 & 216 & 186 & 138 & 54 & 21 & 3 & 0 \\ \hline
A$_{1u}$ & 5 & 19 & 13 & 11 & 2 & 0 & 0 & 0 \\ \hline
A$_{2u}$ & 17 & 20 & 27 & 15 & 10 & 2 & 1 & 0 \\ \hline
E$_{u}$ & 36 & 78 & 84 & 48 & 20 & 6 & 0 & 0 \\ \hline
T$_{1u}$ & 105 & 180 & 219 & 123 & 66 & 15 & 6 & 0 \\ \hline
T$_{2u}$ & 69 & 186 & 168 & 111 & 42 & 12 & 0 & 0 \\ 
\hline
\end{tabular}
\end{center}
\end{table}

We have set up the Hamiltonian matrix in each of the subspaces and obtained 
all the eigenvalues. We have also used a constant $M_{S}$ basis and using the 
full cubic symmetry, factored the space into various irreducible 
representations and obtained all the eigenvalues in each subspace. From the 
eigenvectors, we have computed the total spin of the state. We find a one to 
one correspondence to numerical accuracy, between the two sets of 
calculations. We have also fitted the $\chi T$ vs. T experimental plot by 
using the full spectrum of the Heisenberg Hamiltonian and 
computing \cite{kahn}

\begin{eqnarray}
\chi T=\frac{3}{8}\left[\frac{g^{2}F(J,T)}{1-{zJ^{'}F(J,T)}/{k_{B}T}}\right]
\end{eqnarray}

\noindent where, we have taken the g factor to be 2.1, the ferromagnetic 
exchange constant J to be 27.2 cm$^{-1}$. Here, $\chi T$ is in units of  
$N\mu_B$. The function F(J,T) is given by,

\begin{eqnarray}
F(J,T)=\frac{\sum_{S}\sum_{M_{S}}M_{S}^{2}exp[-E_{0}
(S,M_{S})/k_{B}T]}{\sum_{S}\sum_{M_{S}}exp[-E_{0}(S,M_{S})/k_{B}T]}
\end{eqnarray}

\noindent with $E_{0}(S,M_{S})$ being the eigenvalue of the sum of exchange 
Hamiltonian 
and the magnetic anisotropy term $DS_{Z}^{2}$ and $zJ^{'}$ is the intermolecular 
exchange interaction. Here we have assumed that the molecular anisotropy is 
along the global 
z-axis, and this term is treated as a perturbation to the exchange Hamiltonian 
in Eq. 1. 
In Fig. 4, we show the fit of the experimental data to the model. 

\begin{figure}
\begin{center}
{\includegraphics[width=10.0cm]{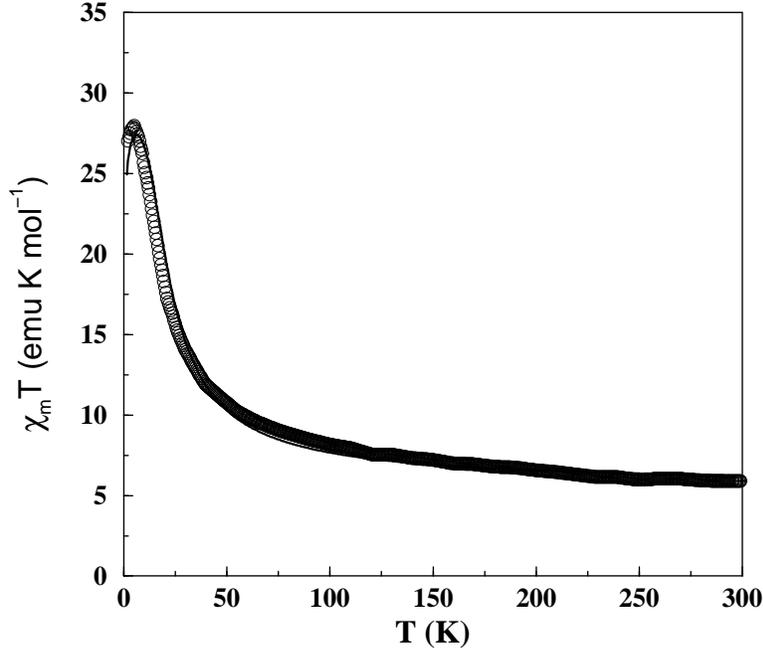}}
\caption{Fit of the $\chi T$ vs. T plot for the $Cu_{6}^{II}
Fe_{8}^{III}$ cluster. The best fit parameters are, J = 27.2 cm$^{-1}$ 
(ferromagnetic), $zJ^\prime$ = -0.008cm$^{-1}$ (antiferromagnetic), 
D = -0.15cm$^{-1}$ and g = 2.1. }
\end{center}
\end{figure}

\section{Summary and Outlook} 
The problem of exploiting total spin invariance together with spatial 
symmetries, especially of non-Abelian point groups has been a long 
standing one.  While full spin symmetry adaptation can be achieved 
by various techniques such as the use of permutation groups, unitary 
groups and the VB method, the last mentioned is the easiest and 
provides easy chemical visualization of the basis states. The main 
objection to the VB technique had been that the basis is nonorthogonal 
and leads to nonsymmetric representation of the Hamiltonian matrix. 
However, with the modification of the Davidson$^,$s algorithm \cite{dav} 
to nonsymmetric matrices by Rettrup \cite{rett}, this objection has ceased 
to be important. The ease with which VB states with any given total spin 
can be generated from objects with assorted individual spins, is 
an advantage which far outweighs the other historical objections to 
the VB method. 

However, even the VB basis suffers from the serious disadvantage like
all other spin adapted methods, when the question of full spatial 
symmetry adaptation comes up. The constant $M_S$ basis methods do 
not suffer from this disadvantage. In this paper, we have demonstrated,
how by combining the ease of spin symmetry adaptation of the VB method
with the spatial symmetry exploitation of the constant $M_S$ methods,
we can devise a scheme which is fully spin and spatial symmetry adapted.
This has been possible because of the ease of transformation of the
VB basis to the constant $M_{S}$ basis. We have demonstrated the power
of the method by applying it to the exchange Hamiltonian of the molecular 
magnet Cu$_6$Fe$_8$ which has cubic symmetry. We note that the hybrid 
VB-constant $M_S$ method also allows easy manipulation of the eigenstates 
of the Hamiltonian for computing properties. The method described here 
can easily be extended to fermionic systems and should provide a 
significant improvement for obtaining exact eigenstates of 
spin conserving model Hamiltonians.

\noindent {\bf Acknowledgement} 
This work was supported through project No. 3108-3
funded by the Indo-French Centre for Promotion of Advanced Research (IFCPAR) 
and by the Swedish Research Council through their grant No. 348-2006-6666.

\end{document}